\documentclass{article}
\usepackage{amsmath,amsthm}
\usepackage{graphicx,epstopdf,epsfig,multirow,epic,bm}
\usepackage{lineno,hyperref}
\usepackage{amssymb}

\normalsize \rm
\begin{document}

\begin{center}
{\Large  \textbf {Assortativity in networks}}\\[12pt]
{\large Fei Ma$^{a,}$\footnote{~The author's E-mail: feima@nwpu.edu.cn. } }\\[6pt]
{\footnotesize $^{a}$  School of Computer Science, Northwestern Polytechnical University, Xi'an 710072, China}\\[12pt]
\end{center}

\begin{quote}
\textbf{Abstract:} The degree-degree correlation is crucial in understanding the structural properties of and dynamics occurring upon network, and is often measured by the assortativity coefficient $r$. In this paper, we first study this measure in detail and conclude that $r$ belongs to an asymmetric range $[-1,1)$ rather than the widely-cited $[-1,1]$. Among which, we verify that star is the unique tree network that achieves the lower bound of index $r$. Next, we obtain that all the resultant networks based on several widely-used kinds of edge-based iterative operations are disassortative if seed model has negative $r$, and also generate a family of growing neutral networks. Then, we propose an edge-based iterative operation to construct growing assortative network when seed is assortative, and further extend it to work well in general setting. Lastly, we establish a sufficient condition for existence of neutral tree network, accordingly, not only find out a representative of any order neutral tree network for the first time, but also are the first to create growing neutral tree networks as well. Also, we obtain $8n/9$ neutral non-tree graphs of distinct order as $n\rightarrow\infty$.\\

\textbf{Keywords:} Degree-degree correlation, Assortativity coefficient, Graph operation, Neutral network. \\

\end{quote}

\vskip 1cm

\section{Introduction}

Complex networks, as a powerful tool for describing complex systems, have been successfully applied in various fields ranging from applied mathematics, statistic physics, computer science to social science \cite{Dorogovtsev-2022,Newman-2018}. As a result, some ubiquitous characteristics on complex systems have been uncovered, including small-world property \cite{Watts-1998}, scale-free feature \cite{Albert-1999}, and so forth. In the literature, it is believed that emergence of these characteristics are often caused by distinct connection patterns between vertices in network \cite{Barabasi-2016}. Among which, the degree-degree correlation is an important pattern, and plays a crucial role in understanding the structural properties of and dynamics occurring upon network \cite{Newman-2002,Menche-2010,Johnson-2010}. 

In fact, some different indices have been proposed in order to characterize the degree-degree correlation \cite{Noldus-2015}. The widely-used one is assortativity coefficient $r$ firstly introduced by Newman \cite{Newman-2002}. Given a network (a.k.a., graph hereinafter) $G=(V,E)$ where $V$ and $E$ represent, respectively, set of vertices and set of edges, the concrete expression of coefficient $r$ is given by

\begin{equation}\label{eqa:MF-PRL-0}
r=\frac{|E|^{-1}\sum\limits_{e_{uv}\in E} k_{u}k_{v}-\left[|E|^{-1}\sum\limits_{e_{uv}\in E} \frac{1}{2}(k_{u}+k_{v})\right]^{2}}{|E|^{-1}\sum\limits_{e_{uv}\in E} \frac{1}{2}(k^{2}_{u}+k^{2}_{v})-\left[|E|^{-1}\sum\limits_{e_{uv}\in E} \frac{1}{2}(k_{u}+k_{v})\right]^{2}},
\end{equation}
in which $e_{uv}$ indicates an edge connecting vertex $u$ to $v$, $k_{u}$ is degree of vertex $u$, and $|X|$ represents the cardinality of set $X$. In the literature, it is a convention that network is considered assortative if $r$ is larger than $0$, and disassortative network corresponds to negative $r$. Clearly, $r=0$ is the critical point that suggests that network under consideration is neutral. Based on this, Newman has shown that many social networks are assortative, however, technological and biological networks have dissortativity \cite{Newman-2002}. In addition, it is believed that parameter $r$ belongs to range $[-1,1]$, i.e., $-1\leq r\leq1$ \cite{Newman-2002,Noldus-2015}. It should be mentioned that network $G=(V,E)$ discussed herein is simple and connected as tried in the literature \cite{Bondy-2008}. We use $G$ as shorthand of $G=(V,E)$ because it is clear from the context.   
    
This paper aims at studying parameter $r$ in more detail, and then reports some new findings.

\section{Main results}
Here, we will discuss assortativity coefficient $r$ in network in more detail. First of all, let us focus on the value for parameter $r$. 

\subsection{Extremal value of assortativity coefficient}
For convenience, we introduce the following expressions 

$$\Gamma(1):=\sum_{e_{uv}\in E}k_{u}\times k_{v},\qquad \Gamma(2):=\sum_{e_{uv}\in E}\left(k_{u}+k_{v}\right), \quad \text{and} \quad \Gamma(3):=\sum_{e_{uv}\in E}\left(k_{u}^{2}+k_{v}^{2}\right),$$
and rewrite 

\begin{equation}\label{eqa:MF-PRL-1}
r=\frac{4|E|\Gamma(1)-\Gamma(2)^{2}}{2|E|\Gamma(3)-\Gamma(2)^{2}}.
\end{equation}

Given a network $G$, using Cauchy-Schwartz inequality, we obtain 

\begin{equation}\label{eqa:MF-PRL-2}
\begin{aligned}\Gamma(2)^{2}&\leq|E|\sum_{e_{uv}\in E}\left(k_{u}+k_{v}\right)^{2}. 
\end{aligned}
\end{equation}
Equality is attained if and only if for all edges $e_{u_{i}v_{i}}$ in network $G$, the next expression holds  

$$\frac{1}{k_{u_{1}}+k_{v_{1}}}=\frac{1}{k_{u_{2}}+k_{v_{2}}}=\cdots=\frac{1}{k_{u_{m}}+k_{v_{m}}},$$
where $m=|E|$. 

Next, we have 

\begin{equation}\label{eqa:MF-PRL-3}
\begin{aligned}2|E|\Gamma(3)-\Gamma(2)^{2}&\geq2|E|\sum_{e_{uv}\in E}\left(k_{u}^{2}+k_{v}^{2}\right)-|E|\sum_{e_{uv}\in E}\left(k_{u}+k_{v}\right)^{2}\\
&=|E|\sum_{e_{uv}\in E}\left(k_{u}-k_{v}\right)^{2}\\
&\geq0.
\end{aligned}
\end{equation}
Clearly, Eq.(\ref{eqa:MF-PRL-3}) achieves minimal value $0$ only if network $G$ is regular graph. That is to say, each vertex in network $G$ has the same degree in this scenario. 

Analogously, we reach to the next expression 

\begin{equation}\label{eqa:MF-PRL-4}
\begin{aligned}2|E|\Gamma(1)+|E|\Gamma(3)&=|E|\sum_{e_{uv}\in E}\left(k_{u}+k_{v}\right)^{2}\\
&\geq\Gamma(2)^{2}.
\end{aligned}
\end{equation}
Equality holds true if and only if network $G$ is either regular graph or balance bipartite graph \cite{MF-1}. If network $G$ is star (a simple example balanced bipartite graph) then 

$$\Gamma(2)^{2}-4|E|\Gamma(1)=2|E|\Gamma(3)-\Gamma(2)^{2},$$ 
which suggests that parameter $r$ is exactly equal to $-1$. To make further progress, we conclude that star is the unique topological structure for tree network that exactly achieves the lower bound of index $r$. Intuitively, in this setting, vertex with highest degree
is completely preferably connected with ones with lowest degree. This is in complete agreement with the concept of disassortative mixing of network. In the other case (i.e., network $G$ is regular graph), we obtain $\Gamma(2)^{2}=4|E|\Gamma(1)$ and then $r=0/0$. This means that parameter $r$ defined in Eq.(\ref{eqa:MF-PRL-0}) is a meaningless index for regular graph. It is worth noting that this viewpoint is also consolidated in the rest of this work.     

\begin{figure}
\includegraphics[width=1\linewidth]{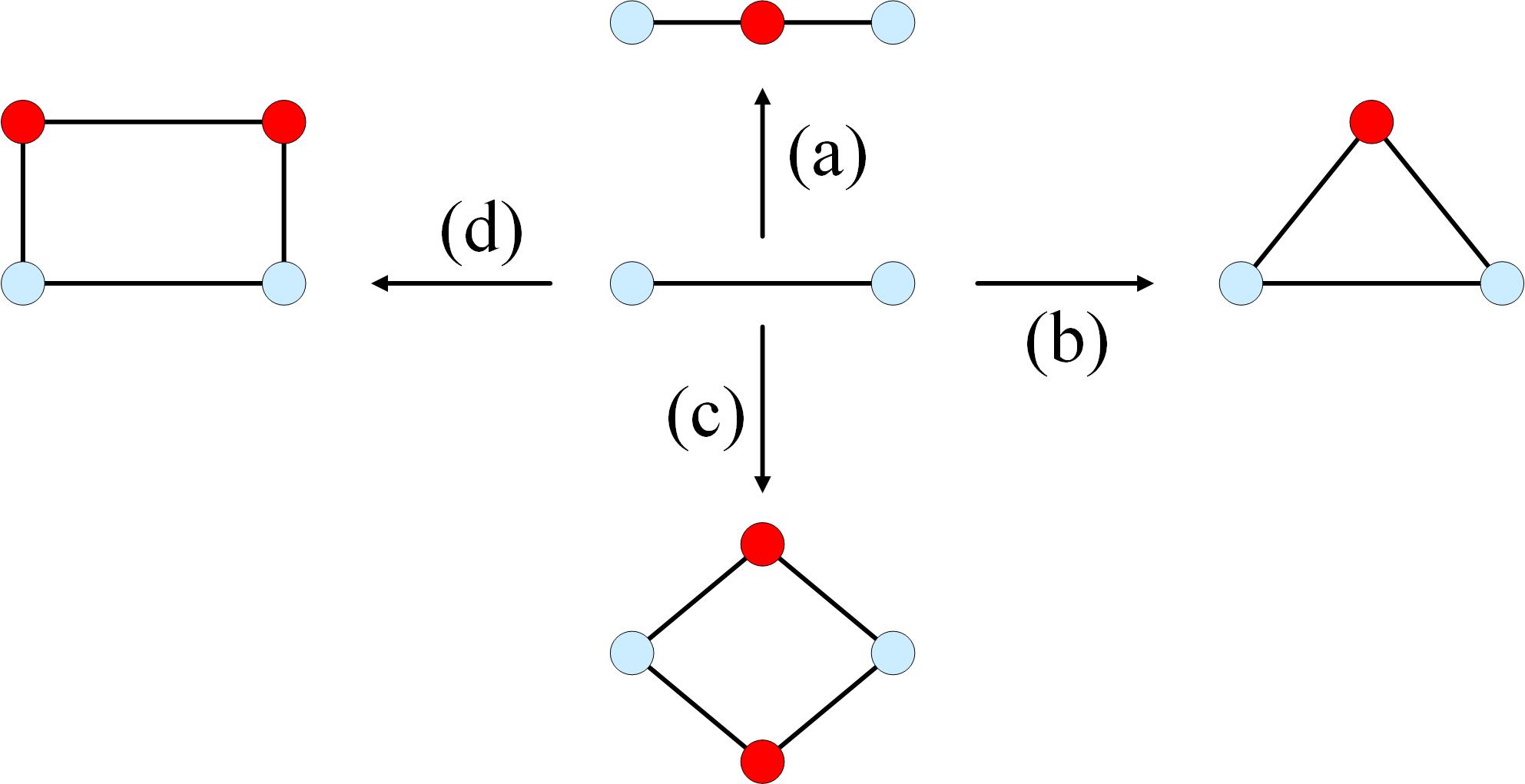}% Here is how to import EPS art
\caption{\label{fig:epsart}(Color online)  The diagram of several edge-based iterative operations including (a) subdivision-operation, (b) triangle-operation, (c) diamond-operation, and (d) rectangle-operation. Concretely speaking,  at each time step, each existing edge is updated with graph at the end of the arrow when creating the corresponding class of growing networks.}
\end{figure}

From Eq.(\ref{eqa:MF-PRL-3}), it follows that if there exist networks with parameter $r=1$, it is clear to the eye that $2\Gamma(1)=\Gamma(3)$ must hold. By definition, it is not hard to check 

$$\Gamma(3)-2\Gamma(1)=\sum_{e_{uv}\in E}(k_{u}-k_{v})^{2}.$$ 
Hence, regular graph is uniquely desirable network that we are seeking for. As mentioned above, however, the assortativity coefficient of regular graph have not yet been properly defined to date. 

Armed with the consequence above, we firmly declare that assortativity coefficient $r$ defined in Eq.(\ref{eqa:MF-PRL-0}) falls into range $[-1,1)$ rather than $[-1,1]$. The range is obviously asymmetric. Naturally, we want to ask what the tight upper hound of parameter $r$ is. The task is left for future research work.

\subsection{Assortativity in graph via simple growth operations}
Now, we discuss about some edge-based iterative operations widely-used to create growing models $G(t)$ for complex network where $t$ represents time step, including subdivision-operation \cite{Ma-2022}, triangle-operation \cite{Dorogovtsev-2002}, diamond-operation \cite{Diggans-2020}, rectangle-operation \cite{Ma-2020}, and determine how they affect the degree-degree correlation of the resulting network. Fig.1 shows these operations. To this end, it is sufficient to study two successive generations of growing network. Without loss of generality, we only consider the original model $G(0)$ (at time step $t=0$) and the first generation $G(1)$ (at time step $t=1$) in the following analysis. Unless otherwise specified, both regular graph and balance bipartite graph are not included in the subsequent discussions.

As the first example operation, subdivision-operation is to insert one vertex on each existing edge of growing network at each time step. For brevity, we denote by $G^{(1)}(t)$ the resulting network after $t$ time steps. Given an arbitrary graph $G$ as original model, we obtain that

$$\Gamma^{(1)}(1):=\sum_{e_{ij}\in E^{(1)}(1)}k_{i}\times k_{j}=2\sum_{e_{uv}\in E}(k_{u}+ k_{v})$$

$$\Gamma^{(1)}(2):=\sum_{e_{ij}\in E^{(1)}(1)}\left(k_{i}+k_{j}\right)=\sum_{e_{uv}\in E}(k_{u}+ k_{v}+4)$$
in which notation $E^{(1)}(1)$ represents edge set of network $G^{(1)}(1)$. Then, it is easy to check

\begin{equation}\label{eqa:MF-PRL-5}
\begin{aligned}4\times|E^{(1)}(1)|\Gamma^{(1)}(1)-\left[\Gamma^{(1)}(2)\right]^{2}&=16|E|\sum_{e_{uv}\in E}(k_{u}+ k_{v})-\left[\sum_{e_{uv}\in E}(k_{u}+ k_{v}+4)\right]^{2}\\
&=-\left[\sum_{e_{uv}\in E}(k_{u}+ k_{v}-4)\right]^{2}\\
&\leq0.
\end{aligned}
\end{equation}
This means that network $G^{(1)}(t)$ can not be assortative regardless of whether original model $G$ is assortative or not. That is to say, subdivision-operation has no positive influence on degree-degree correlation of growing network. In general, the similar statement holds when inserting arbitrarily fixed number $s$ of vertices on each pre-existing edge to construct growing network. Note that this general operation is defined as $s$-order subdivision-operation in the jargon of graph theory \cite{Bondy-2008}. Clearly, case $s=1$ is the subdivision-operation discussed above. It should be noticed that one of greatly potential applications based on this kind of operations is shown at the end of this work.  

Triangle-operation is to update each existing edge in growing network using a triangle at each time step. The famous example network is Pseudofractal scale-free web due to Dorogovtsev \emph{et al} \cite{Dorogovtsev-2002}. As above, we make use of $G^{\bigtriangleup}(t)$ to indicate the $t$-th generation. After that, we have 

$$\Gamma^{\bigtriangleup}(1):=\sum_{e_{ij}\in E^{\bigtriangleup}(1)}k_{i}\times k_{j}=4\sum_{e_{uv}\in E}(k_{u}k_{v}+k_{u}+k_{v}),$$

$$\Gamma^{\bigtriangleup}(2):=\sum_{e_{ij}\in E^{\bigtriangleup}(1)}\left(k_{i}+k_{j}\right)=4\sum_{e_{uv}\in E}(k_{u}+ k_{v}+1),$$
where $E^{\bigtriangleup}(1)$ indicates set of edges in network $G^{\bigtriangleup}(1)$, and obtain

\begin{equation}\label{eqa:MF-PRL-6}
\begin{aligned}&4\times |E^{\bigtriangleup}(1)|\Gamma^{\bigtriangleup}(1)-\left[\Gamma^{\bigtriangleup}(2)\right]^{2}\\
&=-16\left[\sum_{e_{uv}\in E}(k_{u}+ k_{v})\right]^{2}+48|E|\sum_{e_{uv}\in E}k_{u}k_{v}-16|E|^{2}+16|E|\sum_{e_{uv}\in E}(k_{u}+ k_{v})\\
&:=\Gamma^{\bigtriangleup}(4). 
\end{aligned}
\end{equation}
If we select disassortative graph $G$ as original model, quantity $\Gamma^{\bigtriangleup}(4)$ is smaller than zero. This implies that in this setting, the end network $G^{\bigtriangleup}(t)$ is disassortative. We see that the influence from triangle-operation on degree-degree correlation of growing network is greatly dependent on the choice of original model. 

For a growing network, diamond-operation is to replace every existing edge with a pair of paths of length $2$ at each time step. For our purpose, the $t$-th generation is denoted by $G^{\lozenge}(t)$. For network $G^{\lozenge}(t)$, we use notation $E^{\lozenge}(t)$ to stand for the associated edge set, and then write

$$\Gamma^{\lozenge}(1):=\sum_{e_{ij}\in E^{\lozenge}(1)}k_{i}\times k_{j}=8\sum_{e_{uv}\in E}(k_{u}+k_{v}),$$

$$\Gamma^{\lozenge}(2):=\sum_{e_{ij}\in E^{\lozenge}(1)}\left(k_{i}+k_{j}\right)=4\sum_{e_{uv}\in E}(k_{u}+ k_{v}+2).$$
Then, it is not hard to check 

\begin{equation}\label{eqa:MF-PRL-7}
\begin{aligned}4\times|E^{\lozenge}(1)|\Gamma^{\lozenge}(1)-\left[\Gamma^{\lozenge}(2)\right]^{2}&=128|E|\sum_{e_{uv}\in E}(k_{u}+k_{v})-\left[4\sum_{e_{uv}\in E}(k_{u}+ k_{v}+2)\right]^{2}\\
&=-\left[4\sum_{e_{uv}\in E}(k_{u}+ k_{v}-2)\right]^{2}\\
&\leq0.
\end{aligned}
\end{equation}
Based on this, we firmly state that network $G^{\lozenge}(t)$ must be disassortative regardless of whether original model $G$ is assortative or not. That is to say, diamond-operation has a negative influence on degree-degree correlation of growing network. In fact, the similar conclusion holds for more general case, namely, two arbitrary paths serving as candidates. It is just required that the lengths of the chose paths not be smaller than $2$. 

When using rectangle-operation to generate growing network, one needs to update each existing edge by rectangle at every time step. After $t$ time steps, the resultant network is denoted by notation $G^{\Box}(t)$. Accordingly, $E^{\Box}(t)$ indicates edge set. As before, we gain 

$$\Gamma^{\square}(1):=\sum_{e_{ij}\in E^{\square}(1)}k_{i}\times k_{j}=4\sum_{e_{uv}\in E}(k_{u}k_{v}+k_{u}+k_{v}+1),$$

$$\Gamma^{\square}(2):=\sum_{e_{ij}\in E^{\square}(1)}\left(k_{i}+k_{j}\right)=4\sum_{e_{uv}\in E}(k_{u}+ k_{v}+2),$$
and 
 
\begin{equation}\label{eqa:MF-PRL-8}
\begin{aligned}4\times|E^{\square}(1)|\Gamma^{\square}(1)-\left[\Gamma^{\square}(2)\right]^{2}&=64|E|\sum_{e_{uv}\in E}(k_{u}k_{v}+k_{u}+k_{v}+1)-\left[4\sum_{e_{uv}\in E}(k_{u}+ k_{v}+2)\right]^{2}\\
&=16\left\{4|E|\sum_{e_{uv}\in E}k_{u}k_{v}-\left[\sum_{e_{uv}\in E}(k_{u}+ k_{v})\right]^{2}\right\}\\
&=16[4|E|\Gamma(1)-\Gamma^{2}(2)].
\end{aligned}
\end{equation}
It means that assortativity coefficient of growing network $G^{\Box}(t)$ has the same sign as that of original model $G$. 

According to Eq.(\ref{eqa:MF-PRL-8}), given an arbitrary regular graph as original model, it is not difficult to prove that the resulting network $G'(t)$ has zero assortativity coefficient, and hence is neutral. It is worth mentioning that parameter $r$ of network $G'(t)$ is exactly equal to $0$. To the best of our knowledge, this is the first graph that achieves the critical point of parameter $r$ completely. It should be mentioned that after obtaining graph $G'(t)$, one can obtain a family of neutral networks by rewiring strategy only if degree sequence keeps unchanged. In addition, it seems reasonable that regular graph is deemed as neutral network.  

\begin{figure}
\includegraphics[width=1\linewidth]{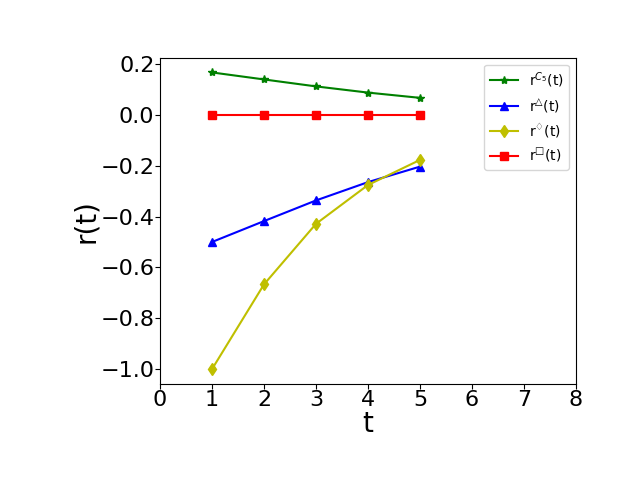}% Here is how to import EPS art
\caption{\label{fig:epsart}(Color online)  The diagram of assortativity coefficient of growing networks $G^{\bigtriangleup}(t)$, $G^{\lozenge}(t)$, $G^{\Box}(t)$ and $G^{C_{5}}(t)$. Note that $4$-cycle is selected as original model in all the settings. Obviously, networks $G^{\bigtriangleup}(t)$ and $G^{\lozenge}(t)$ are disassortative, $G^{\Box}(t)$ is neutral, and $G^{C_{5}}(t)$ is assortative. So, numerical simulations are in line with theoretical analysis.}
\end{figure}

In a nutshell, if a given graph $G$ as original model is disassortative, so are all the end networks obtained from these four operations mentioned above. See Fig.2 for illustrative examples. It is natural to ask whether there is or not one edge-based iterative operation for creating growing assortative network given an arbitrary graph serving as original model. In order to answer this question, let us introduce a new operation, i.e., $5$-cycle operation ($C_{5}$-operation for short). Concretely speaking, $C_{5}$-operation is to update each existing edge by $C_{5}$ at every time step. After $t$ time steps, we obtain the resultant network $G^{C_{5}}(t)$. By analogy with analysis shown in Eq.(\ref{eqa:MF-PRL-8}), we prove that if a assortative graph $G$ serves as original model, the resultant growing network $G^{C_{5}}(t)$ must have assortative mixing structure. To make further progress, it is not difficult to verify that $n$-cycle operation is a candidate for constructing growing assortative network. Note that parameter $n$ is no less than $5$ and also depends on the size of original graph.

\begin{figure}
\includegraphics[width=0.7\linewidth]{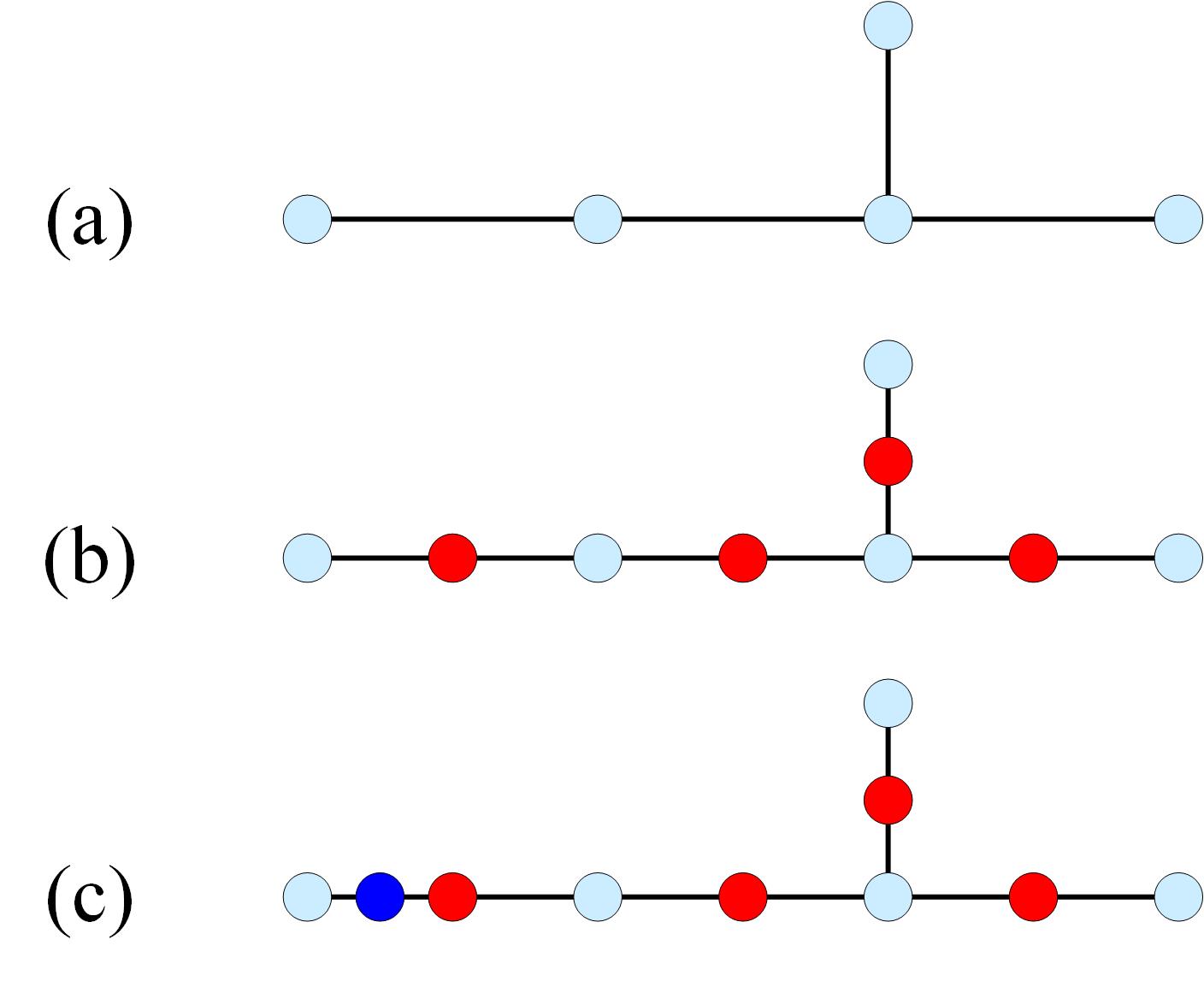}% Here is how to import EPS art
\caption{\label{fig:epsart}(Color online) The diagram of neutral tree networks. Tree network in panel (b) is obtained from tree network in panel (a) by manipulating subdivision-operation, and then turns out to be neutral. The added vertices are highlighted in red. This further suggests that there exists an effective approach to create a family of growing neutral tree networks. Now, one vertex highlighted in blue is inserted into an edge of tree in (b), which yields a new tree. An example is shown in panel (c). It is easy to verify that the resulting tree has zero assortativity coefficient, and is also neutral.}
\end{figure}

\subsection{Representative of neutral network}

Below, we will focus on neutral network. First, we study tree network. Then, non-tree network is discussed.

Let us now be concerned on assortativity coefficient of tree network. Before diving into detailed discussions, we need to rethink of Eq.(\ref{eqa:MF-PRL-5}). After that, it is clear to see that for a given graph $G$, if the following equation holds

\begin{equation}\label{eqa:MF-PRL-9}
\sum_{e_{uv}\in E}(k_{u}+ k_{v})=4|E|,
\end{equation}
the resulting graph $G^{(1)}(t)$ must be neutral. In addition, it is easy to prove that an arbitrary cycle satisfies Eq.(\ref{eqa:MF-PRL-9}). From which, we can easily create a tree having $2n+1$ vertices that turns out to be neutral. Concretely speaking, we execute the following procedure: (1) deleting an arbitrary edge in given cycle of length $n(\geq3)$, (2) connecting a new vertex to any vertex of degree two, and (3) performing subdivision-operation on the end tree in (2). Obviously, the resulting tree network is also subject to Eq.(\ref{eqa:MF-PRL-9}). Panel (b) in Fig.3 shows an example. In fact, the procedure above can be extended to obtain many desirable tree networks. Specifically, we are given two paths in which at least one has length no less than $3$, then identify one leaf vertex of one path and any degree $2$ vertex in the other path into one vertex. See panel (a) in Fig.3 for an example in which an edge and a length $3$ path are chose as candidates. In a word, Eq.(\ref{eqa:MF-PRL-9}) is a sufficient condition for the existence of the neutral feature of tree network. 

Furthermore, we are able to obtain neutral tree network having number $n+4$ of vertices using the procedure above. Specifically, it is sufficient to insert proper number of vertices on an arbitrarily selected edge in the resultant tree in (3). See panel (c) in Fig.3 for an illustrative example. In general, the choice of candidate edges in the resulting tree in (3) is not necessary an edge but any number of edges. Based on this, we prove for the first time that there exist neutral tree networks with no less than $7$ vertices. Additionally, it is easy to show that an arbitrary tree is not assortative when its number of vertices is no more than $6$. To sum up, we have the ability to find out a representative of any order neutral tree network.

Following the discussions above, it is natural to propose an algorithm available for yielding growing neutral tree networks, as follows: 

\begin{itemize}
\item At $t=0$, the seminal model is an arbitrary tree subjected to Eq.(\ref{eqa:MF-PRL-9}), denoted by $T(0)$.

\item At $t\geq1$, tree network $T(t)$ is created from the preceding model $T(t-1)$ by implementing $p_{t}$-order subdivision-operation.

\end{itemize}
Notice that $p_{t}$ is an arbitrary positive integer. An illustrative example is displayed in panel (b) of Fig.3. To our knowledge, model $T(t)$ is the first growing neutral tree network. The tendency of assortativity coefficient $r_{t}$ of some tree networks $T(t)$ is plotted in Fig.4. Clearly, numerical simulations are perfectly consistent with theoretical analysis. 

\begin{figure}
\includegraphics[width=1\linewidth]{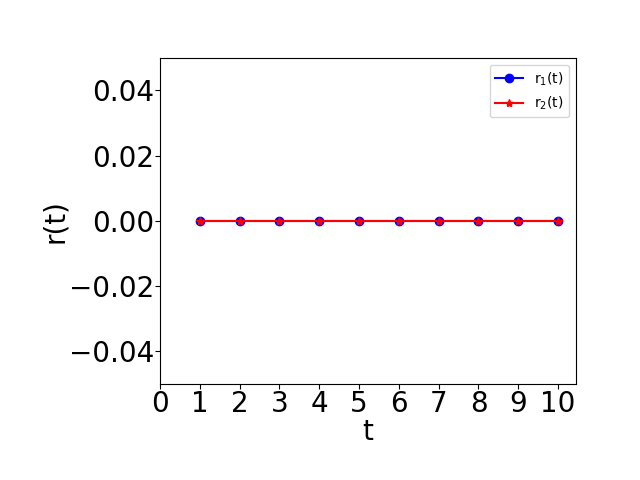}% Here is how to import EPS art
\caption{\label{fig:epsart}(Color online)  The diagram of assortativity coefficient $r_{i}(t)$ of two growing tree networks $T_{i}(t)$ where $i=1,2$. Tree in panel (a) of Fig.3 is selected as original model. Parameter $p_{t}\equiv1$ holds in the development of tree $T_{1}(t)$. We assume parameter $p_{t}=t$ in the other growing tree network $T_{2}(t)$. It is clear to see that two tree networks $T_{i}(t)$ have zero assortativity coefficient, and are thus neutral. }
\end{figure}

By far, we succeed in finding a representative of any order neutral graph. It should be mentioned that the representative is a tree. Next, we are concerned with other representatives having non-tree structure    

Given any cycle $C_{n}$, the corresponding graph $C_{n}^{\Box}(1)$ is neutral. Similarly, given any neutral tree $T$, the corresponding graph $T^{\Box}(1)$ is also neutral. Based on this, it is clear to see that there exists neutral non-tree graph of either order $3(n-1)$ or order $3(n-1)+1$. Furthermore, we also obtain that there is neutral non-tree graph of order $f(x,y)$ 

\begin{equation}\label{eqa:MF-PRL-10}
f(x,y):=(2y+1)x-2y, \qquad x,y\in \mathbb{N}, \quad x\geq7, \quad y\geq1.
\end{equation}
Clearly, if $x|[3(n-1)+2]$ and $2y|[3(n-1)+2]$, then $f(x,y)|[3(n-1)+2]$. At the same time, for a given parameter $n$, one is able to find out at least $\lfloor\frac{n+2}{3}\rfloor$ neutral non-tree graph of distinct order. To sum up, one can obtain at least $\lfloor\frac{n+2}{3^{2}}\rfloor$ neutral non-tree graph of distinct order

Taken together, we come to the following statement that given $n\rightarrow\infty$, there are at least $7n/9$ neutral non-tree graphs of distinct order.  Along the line of similar research, we can conclude that there are about $8n/9$ neutral non-tree graphs of distinct order as $n\rightarrow\infty$.

\section{Conclusion}
To conclude, we study the degree-degree correlation in complex network by virtue of assortativity coefficient $r$ in more detail, and obtain that (i) $r$ belongs to an asymmetric range $[-1,1)$ rather than the widely-known $[-1,1]$, (ii) star is the unique topological structure for tree network that achieves the lower bound of index $r$, (iii) a family of growing neutral networks is generated, (iv) a sufficient condition for existence of the neutral feature of tree network is established, (v) a representative of any order neutral tree network is found, (vi) an algorithm suitable for growing neutral tree networks is proposed, and (vii) there are at least $8n/9$ neutral non-tree networks of distinct order as $n\rightarrow\infty$. The results obtained here, particularly those related to neutral networks, are useful to in depth investigate the influence of the degree-degree correlation on structural property on complex network. We will focus on discussion between the degree-degree correlation and dynamics taking place on networks in the future.

\section*{Acknowledgments}
The research was supported by the Fundamental Research Funds for the Central Universities No. G2023KY05105 and the National Key Research and Development Plan under grant 2020YFB1805400.


{\footnotesize
\begin{thebibliography}{9}

\setlength{\parskip}{0pt}


\bibitem{Dorogovtsev-2022} S.N. Dorogovtsev, J.F.F Mendes. The nature of complex networks. Oxford University Press, 2022.

\bibitem{Newman-2018} M.E.J. Newman. Networks. Oxford University Press, 2018.

\bibitem{Watts-1998} D.J. Watts, S.H. Strogatz. Collective dynamics of `small-world' networks. Nature, 1998, 393(6684): 440-442.

\bibitem{Albert-1999} A.-L. Barab\'{a}si, R. Albert. Emergence of scaling in random networks. Science, (1999): 5439, 509-512.

\bibitem{Barabasi-2016} A.-L. Barab\'{a}si. Network Science. Cambridge University Press. 2016.

\bibitem{Menche-2010} J. Menche, A. Valleriani, R. Lipowsky. Asymptotic properties of degree-correlated scale-free networks. Physical Review E, 2010, 81(4): 046103.

\bibitem{Johnson-2010} S. Johnson, J.J. Torres, J. Marro, M.A. Munoz. Entropic origin of disassortativity in complex networks. Physical Review Letters, 2010, 104(10): 108702.

\bibitem{Newman-2002} M.E.J. Newman. Assortative mixing in networks. Physical Review Letters, 2002, 89(20): 208701.

\bibitem{Noldus-2015} R. Noldus, P. Van Mieghem. Assortativity in complex networks. Journal of Complex Networks, 2015, 3(4): 507-542.

\bibitem{Bondy-2008} J. Bondy, U. Murty. Graph Theory. Springer, 2008.

\bibitem{MF-1} A graph $G$ is bipartite if its vertex set $V$ can be partitioned into two subsets $V_{1}$ and $V_{2}$ so that every edge has one end in $V_{1}$ and one end in $V_{2}$. Bipartite graph is considered balanced if each vertex in subset $V_{1}$ has the same degree with each other and similarly for vertices in subset $V_{2}$. 

\bibitem{Ma-2022} F. Ma, P. Wang, X. Luo. A method for geodesic distance on subdivision of trees with arbitrary orders and their applications. IEEE Transactions on Knowledge and Data Engineering, 2022, 34(5): 2063-2075.

\bibitem{Dorogovtsev-2002} S.N. Dorogovtsev, A.V. Goltsev, J.F.F. Mendes. Pseudofractal scale-free web. Physical Review E, 2002, 65(6): 066122.

\bibitem{Diggans-2020} C.T. Diggans, E.M. Bollt, D. Ben-Avraham. Stochastic and mixed flower graphs. Physical Review E, 2020, 101(5): 052315.

\bibitem{Ma-2020} F. Ma, X. Wang, P. Wang. An ensemble of random graphs with identical degree distribution. Chaos: An Interdisciplinary Journal of Nonlinear Science, 2020, 30(1): 013136.




\end{thebibliography}
}
\end{document}